\documentstyle[12pt,a4,epsfig]{article} 

\begin{document}
\unitlength=0.9cm
\title{{\Large\bf Drift- or Fluctuation-Induced Ordering and Self-Organization
in Driven Many-Particle Systems}\\[8mm] 
{\normalsize Dirk Helbing$^{1,2}$ and Tadeusz P{\l}atkowski$^{3,1}$\\[3mm]
$^1$ Institute for Economics and Traffic,
Dresden University of Technology, 01062 Dresden, Germany\\[2mm]
$^2$ CCM-Centro de Ci\^{e}ncias Matem\'{a}ticas, Universidade da Madeira, 
Campus Universit\'{a}rio da Penteada,
9000-390 Funchal, Madeira, Portugal\\[2mm]
$^3$ Dept. of Mathematics, Informatics and Mechanics, 
University of Warsaw, Banacha 2, 02-097 Warsaw, Poland\\ \mbox{ }}}
\maketitle
\begin{abstract}
According to empirical observations, some pattern formation phenomena in driven many-particle systems
are more pronounced in the presence of a certain noise level. We investigate this phenomenon
of fluctuation-driven ordering with a cellular automaton model of interactive motion in space and find
an optimal noise strength, while order breaks down at high(er) fluctuation levels.
Additionally, we discuss the phenomenon of noise- and drift-induced self-organization
in systems that would show disorder in the absence of fluctuations. In the future,
related studies may  have applications to the control of many-particle 
systems such as the efficient separation of particles. The rather general formulation of 
our model in the spirit of game theory may allow to shed some light
on several different kinds of noise-induced ordering phenomena observed in physical, chemical, 
biological, and socio-economic systems (e.g., attractive and repulsive agglomeration, 
or segregation). 
\end{abstract}
\clearpage
Noise-related phenomena can be quite surprising and exciting. Therefore, they
have attracted the interest of many researchers. For example, we 
mention stochastic resonance \cite{stochres}, 
structural (in)stability \cite{Pri,Pri2},
noise-driven motion (`Brownian motors') \cite{biol1a,biol1c,brownmot}, and ``freezing by heating''
\cite{freezing}. The approach proposed in this contribution
differs from these phenomena. Moreover, since both, the
initial conditions and the interaction strengths in our model are 
assumed to be independent of the position in space, 
the fluctuation-induced self-organization
discussed later on should be distinguished from 
so-called ``noise-induced  transitions''  in systems with multiplicative noise as well, 
where we have a space-dependent diffusion coefficient which 
can induce a transition \cite{HoLe84}. Although our model 
is related with diffusive processes, it is also different from reaction-diffusion systems
that can show fluctuation-induced self-organization phenomena known as
Turing patterns \cite{turing,turing1,turing2,turing3,turing4,turing5}.
\par
The main goals of this contribution are (i) to qualitatively understand the observed 
noise-induced ordering phenomena in certain self-driven many-particle systems and (ii) to
derive and investigate a simplified mathematical model for them.
Some of the properties of this model appear reminiscent of driven (e.g.\ shared) heterogeneous
granular systems. For example, one finds stratification phenomena such as the segregation of
different kinds of grains into layers \cite{Sant2ScHe96,MakHaKiSt97}.
This mechanism explains, for example, some of the
geological formations and ore concentrations in the earth. As it may be also used to
separate different materials, it is interesting to ask for the most
efficient separation technique including the optimal noise level. A similar segregation phenomenon
of lane formation has been found in pedestrian 
counter-streams \cite{HelMo295}. This is based on
a reduction of the interaction strength and related with an increase in the efficiency
or average speed of motion \cite{helvic}. 
\par
In the following, we will explore a cellular automaton model which 
extends the one discussed in Ref.~\cite{helvic} by the effects of fluctuations, drift, and
asymmetric interactions. It may, for example, 
reflect the interactive one-dimensional motion of some driven many-particle systems
in a spatial direction perpendicular to the main flow direction(s) and to the boundaries. 
For this purpose, let us imagine the example of pedestrian streams in a corridor 
with two opposite flow directions $a$. 
We subdivide the one-dimensional space into cells $i$ comparable to
the shoulder width (diameter) $\Delta x$ of the pedestrians (particles, entities). Speaking in more
general terms, we have $A$ subpopulations {$a$} with  {$N_a$} entities {$\alpha$}
distributed over {$I$} cells $i\in \{1,\dots,I\}$. We denote the number of entities in cell $i$
belonging to subpopulation $a$ by $n_i^a$. Moreover, we represent the
kind of interaction and the interaction strength between two entities of
subpopulations $a$ and $b$ by a constant parameter value $P_{ab}$. 
(Generalizations are easily possible.) Finally, we update
the locations of the entities $\alpha$
according to the following rules:
{\em 1st step:} Select the entity $\alpha$ randomly. If
$\alpha$ is located in cell $i$ and
belongs to subpopulation $a$,  determine 
\begin{equation}
S_\alpha(j,t)=\sum_{b}P_{ab}\, n_{j}^{b}(t) +\xi _j^{\alpha }(t) 
\label{success}
\end{equation}
for $j=i$ and the nearest neighbors $j= i\pm 1$, where 
$\xi_j^\alpha(t)$ are random fluctuations uniformly distributed in the interval $[-p_a,p_a]$,
so that  $p_a$ denotes the fluctuation strength.
{\em 2nd step:} Move to the neighboring cell $i\pm 1$ with probability 
\begin{equation}
 P_\alpha(i\pm 1|i;t) \propto \max\{0,S_\alpha(i\pm1,t) - S_\alpha(i,t) \} \, .
\label{propim}
\end{equation}
{\em 3rd step:} Repeat steps 1 and 2 until the locations of $N = \sum_a N_a$ 
entities were updated.
{\em 4th step:} With probability $V_a^0$, move all entities $\alpha$ of subpopulation $a$  
by one cell into the same direction.
{\em 5th step:} Return to step 1.
\par
Formula (\ref{success}) calculates the expected effect of interactions with other entities.
$S_a(i,t)$ is a potential function, see Eq.~(\ref{pot}). In the language of
game theory, it can be called the {\em expected success}, 
since, according to the {\em proportional imitation rule} 
(\ref{propim}), an entity $\alpha$ moves to a neighboring cell $i\pm 1$ only if it can increase 
the value of $S_\alpha$. 
The values $P_{ab}$ may be interpreted as {\em payoffs} in interactions between two entities
of subpopulations $a$ and $b$. $P_{ab}$ is positive for attrative, cooperative, or profitable
interactions, while it is negative for repulsive, competitive, or loss-making interactions.
The fourth step reflects a bias in the motion of the particles of subpopulation $a$, i.e.
a {\em drift velocity}.
\par
We have carried out various simulations with random initial and periodic
boundary conditions, in order to have a translation-invariant system. Note that
self-organized pattern formation in such a system implies spontaneous symmetry-breaking
and a pronounced history-dependence of the resulting state. The typical solutions
are dependent on the specified payoffs $P_{ab}$, fluctuation strengths $p_a$, and
drift velocities $V_a^0$. Replacing the asynchronous (random sequential) update of
steps 1 to 3 by a parallel update yields similar (but less random, i.e. ``nicer looking'') results. 
Replacing the parallel update of the velocities in step 4 by an asynchronous one induces such a high
noise level that the system is often disordered. 
\par
\begin{figure}[htbp]
\begin{center}
\begin{picture}(15,6.5)
\put(-0.1,-0.6){\epsfig{width=7.1\unitlength, angle=0,
      file=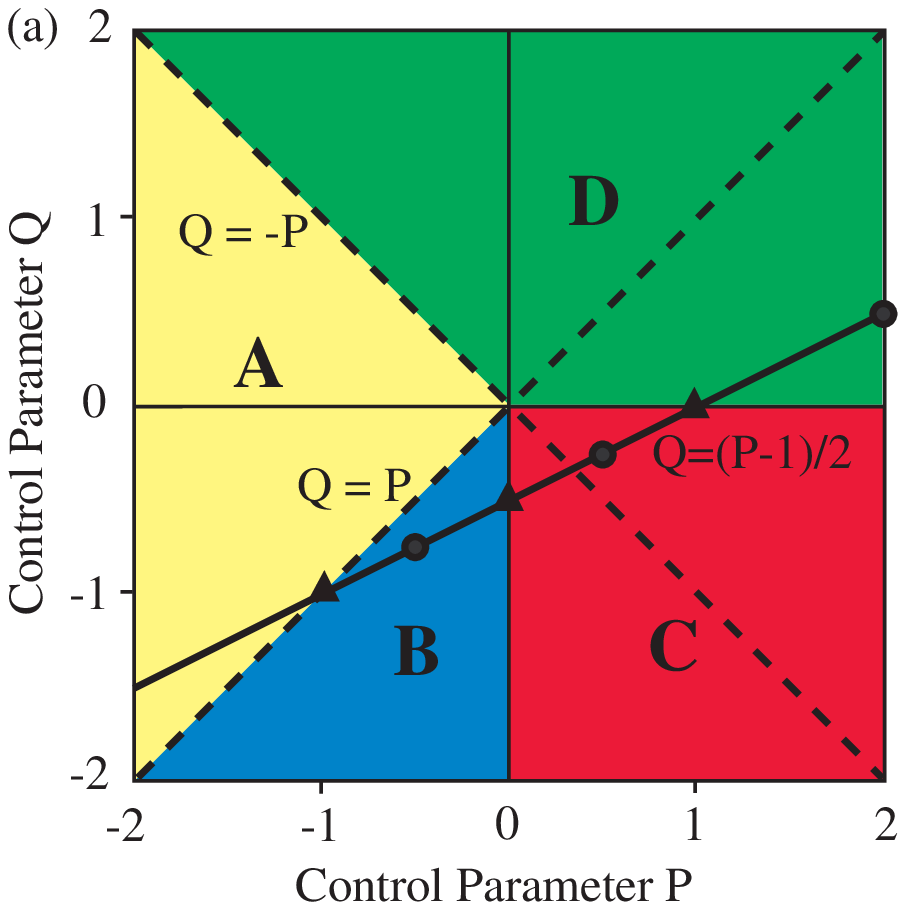}}
\put(6.9,6.7){\epsfig{width=7.5\unitlength, angle=-90,
      file=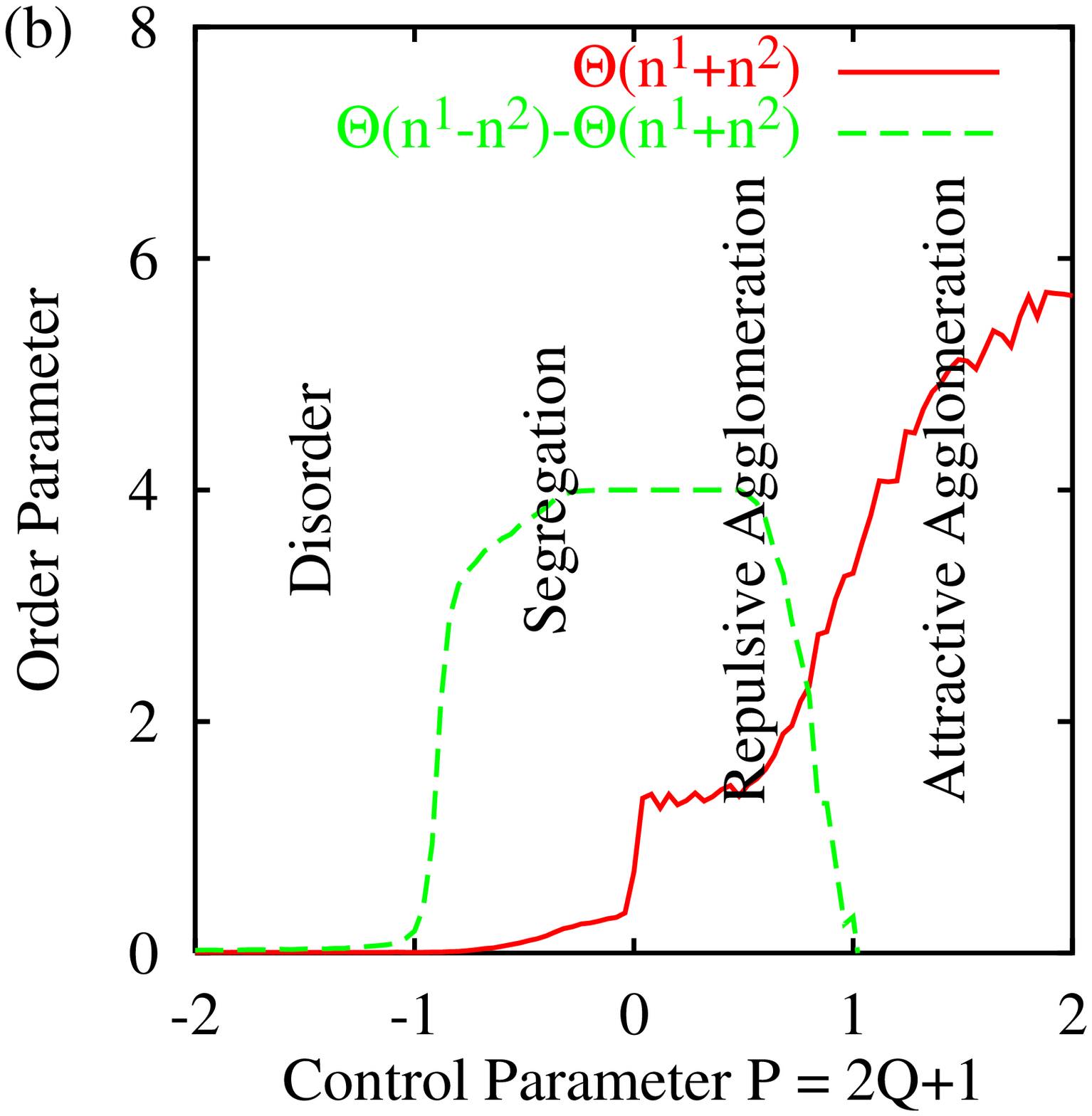}}
\end{picture}
\end{center}
\caption[]{(a) Theoretical phase diagram of the four qualitatively different patterns 
resulting for the symmetric model without diffusion: A = disordered phase,
B = segregation, C = repulsive agglomeration, D = attractive agglomeration.
(b) Order parameters along the line $Q=(P-1)/2$ (see solid line in Fig.~\ref{Fig1}a), 
averaged over 20 runs after a time period of $20\,000(N_1+N_2)$
update steps with $I=200$ cells and $N_1 = N_2 = 2000$ entities in each of two subpopulations. 
The theoretically predicted phase transitions at $P=-1$, $P=0$, and $P=1$ 
(see black triangles in Fig.~\ref{Fig1}a) are clearly visible.
\label{Fig1}}
\end{figure}

To obtain a theoretical understanding of our simulation results, 
we have derived mean value equations for 
the densities $\rho_a(x,t) = n_i^a(t) / \Delta x$ 
with $x = i \Delta x$. For this purpose, we have derived a master equation and determined
the drift- und fluctuation-coefficients in the usual way. By second order Taylor expansion,
the resulting equations can be then approximately
cast into the form of Fokker-Planck equations \cite{QuantSoz}
\begin{equation}
\frac{\partial \rho_a(x,t)}{\partial t} +
\frac{\partial}{\partial x} \left[ \rho_a(x,t)V_a(x,t) \right] 
= D_a \frac{\partial^2 \rho_a(x,t)}{\partial x^2}  
\label{FPG}
\end{equation}
where the diffusion coefficients $D_a$ increase with the 
fluctuation strength $p_a$ in a roughly proportional way. Moreover, $D_a$ vanishes
when $p_a$ vanishes and $|\partial S_a(x,t)/\partial x|$ is small
(as for our homogeneous initial conditions). The exact relation for $D_a$ is rather complicated,
but not of interest, here.  Finally, the drift coefficients are given by
\begin{equation}
V_a(x,t) = V_a^0 + \frac{\partial S_a(x,t)}{\partial x} 
\qquad \mbox{with} \qquad
S_a(x,t) = \sum_b P_{ab} \, \rho_b(x,t) \, .
\label{pot}
\end{equation}
Accordingly, we are confronted with non-linearly coupled Burgers equations, which
may show a diffusion instability.  
In the following, we will check whether
the above mean value equations yield qualitatively meaningful results, 
i.e.\ whether correlations can be neglected. For this purpose, we will carry out a linear stability 
analysis and compare the theoretical phase diagram with the numerically determined
one.  In the case of two subpopulations $a\in \{1,2\}$ and $V_1^0 =V_2^0 = 0$, 
the homogeneous solution $\rho_a^0 = \overline{n_i^a}/\Delta x$ with 
$\overline{n_i^a} = N_a / I$  
should be unstable with respect to fluctuations, if
\begin{equation}
 \rho_1^0 P_{11} + \rho_2^0 P_{22} > D_1 + D_2
\label{cond1}
\end{equation} 
or
\begin{equation}
 \rho_1^0 \rho_2^0 P_{12} P_{21}  >  (\rho_1^0 P_{11} - D_1)(\rho_2^0 P_{22} - D_2) \, .
\label{cond2}
\end{equation}
Let us first discuss the the symmetric
case with $\rho_1^0 = \rho_2^0 = \rho$, $P_{11} = P_{22} = P$, 
$P_{12} = P_{21} = Q$, and vanishing diffusion $D_1 = D_2 = 0$. Then, condition
(\ref{cond1}) reduces to $2 \rho P > 0$, and condition (\ref{cond2}) becomes
$Q^2 >  P^2$. We can distinguish the following solutions
(see Fig.~\ref{Fig1}a, for representatives see Fig.~4 in Ref.~\cite{helvic}): 
A) If $P<0$ and $Q^2 < P^2$ [i.e. $P< 0$ and $P<Q<-P$)],
a {\em homogeneous distribution} $\rho_a(x,t) = \rho_a^0$ over all sites
in {\em both} subpopulations is stable with respect to
small perturbations (which corresponds to {\em disorder}). B) If $P<0$ (self-repulsion) and 
$Q<0$ (repulsion between the subpopulations), but $Q<P$, we should find
{\em segregation} (with a tendency that {\em all} sites are equally occupied, but {\em either} by
one subpopulation {\em or} by the other).
C) If  $Q<0$ (repulsion), but $P>0$ (self-attraction),
we expect {\em repulsive agglomeration} 
(i.e. both subpopulations should cluster at {\em different} sites, with empty sites in between). D)
If $Q>0$ (attraction) and $Q > -P$, we should have {\em attractive agglomeration} 
(clustering of both subpopulations at the {\em same} sites, with empty sites in between). 
Consequently, on the
line $Q = (P-1)/2$ (i.e. for $P=2Q+1$), we should cross the phase boundary between
disorder and segregation at $P=-1$, the one between segregation and repulsive
agglomeration at $P=0$, and the one between repulsive and attractive agglomeration
at $P=+1$. This is, in fact, 
confirmed by our numerical simulations (see Fig.~\ref{Fig1}b),
so that we can trust the instability analysis based on the
mean value equations. The reason for this is the local nature of the interactions.
To characterize the different states, we have used 
order parameters of the form
\begin{equation}
 \Theta(y) = \frac{1}{I} \sum_{i=1}^I (y_i  - \overline{y_i})^2 \qquad
\mbox{with} \qquad \overline{y_i} = \frac{1}{I} \sum_{i=1}^I y_i 
\end{equation}
to measure the variances of 
(i) $y_i= (n_i^1/\overline{n_i^1}+n_i^2/\overline{n_i^2})$
(i.e. the deviation from a homogeneous occupation of {\em all} sites), 
or (ii) $y_i=(n_i^1/\overline{n_i^1} - n_i^2/\overline{n_i^2})$
(i.e. the difference in the degree of occupation by different subpopulations).
$\Theta(n^1+n^2)$ is sensitive to (attractive or repulsive) agglomeration (i.e. to clustering
with empty sites in between), and $\Theta(n^1+n^1)$ recognizes, 
when the two subpopulations tend to use different sites (as for segregation
or repulsive agglomeration).
\par
\begin{figure}[htbp]
\begin{center}
\begin{picture}(16,6.5)
\put(0,6.7){\epsfig{width=7.5\unitlength, angle=-90,
      file=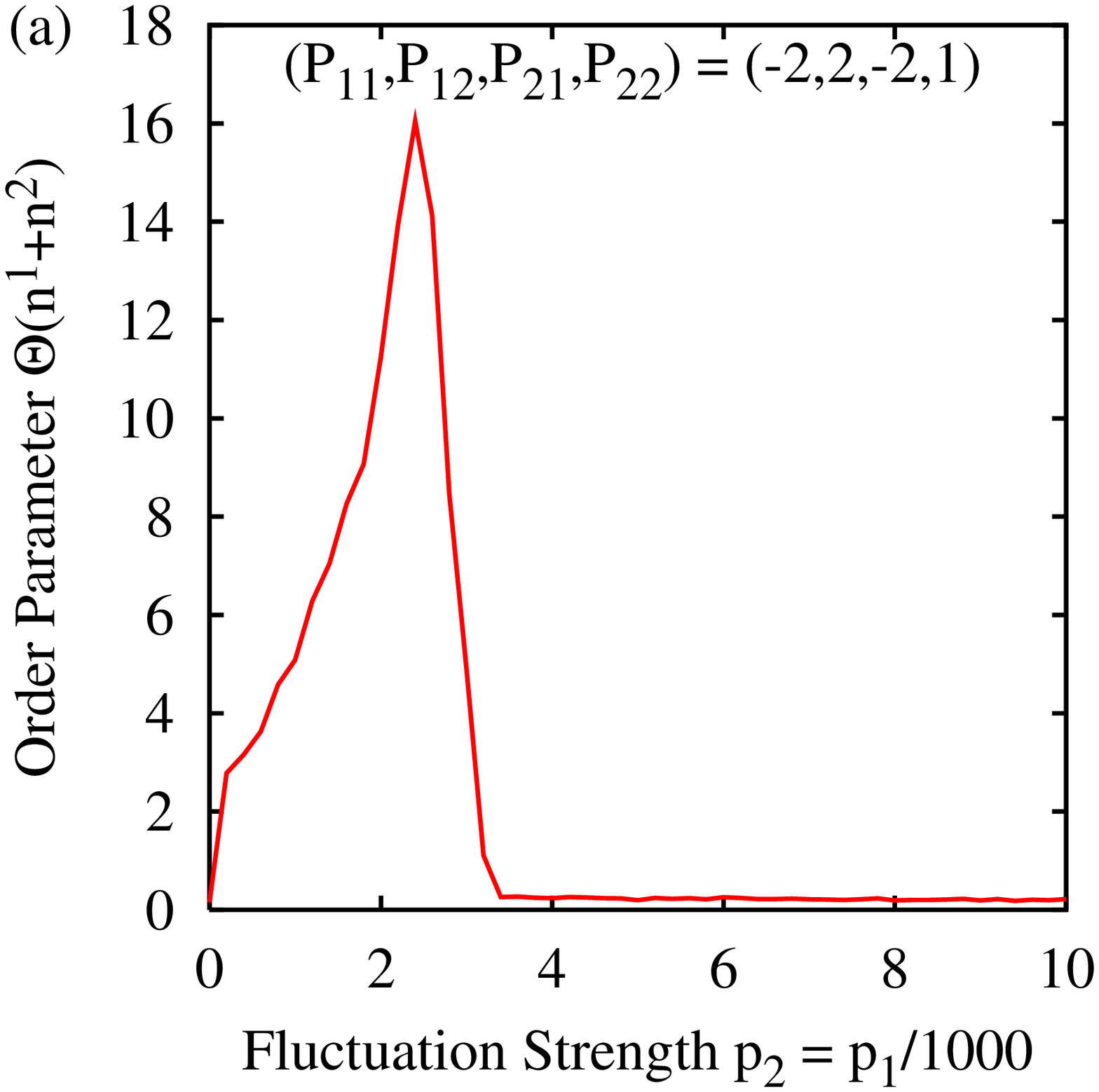}}
\put(7.8,6.7){\epsfig{width=7.5\unitlength, angle=-90,
      file=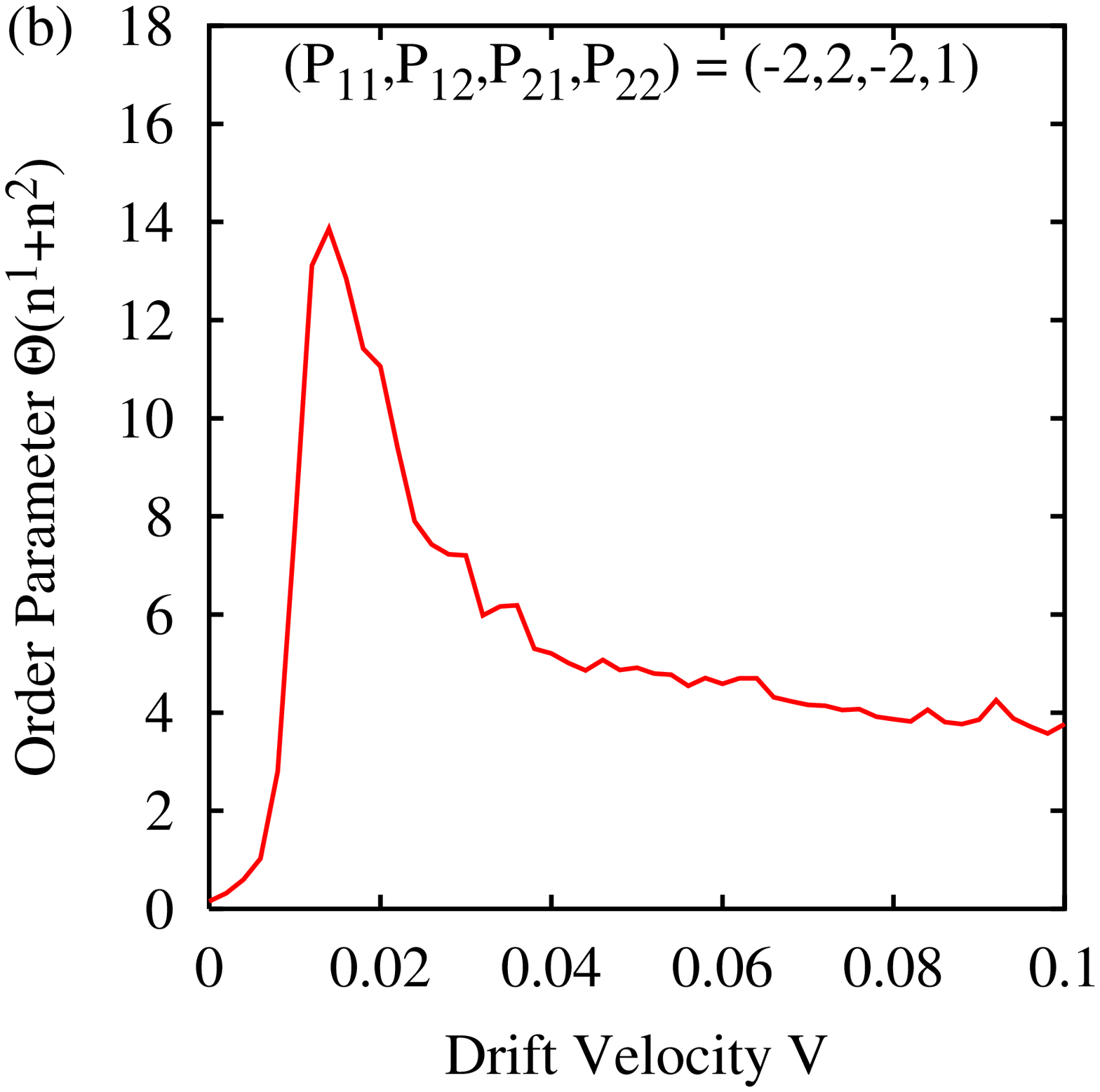}}
\end{picture}
\end{center}
\caption[]{Order parameter for the degree of agglomeration
as a function of (a) the fluctuation strength $p_2 = 0.001 p_1$
(with $V_1^0 = V_2^0 = 0$) and (b) the drift velocity $V = V_1^0$ 
(with $V_2^0 = 0 = p_1 = p_2$) for an example with 
asymmetric interactions. The curves are averages over 20 runs 
after a time period of $20\,000(N_1+N_2)$ update steps for
$I = 20$ cells and $N_1=N_2 = 200$ entities in each subpopulation.\label{Fig2}}
\end{figure}

Let us now focus on the general case with arbitrary payoffs and diffusion. While $D_1, D_2 > 0$
increases the threshold for pattern formation in Eq.~(\ref{cond1}), in Eq.~(\ref{cond2}) 
it can reduce the threshold for moderate diffusion, while the threshold will be higher for
large diffusion. We are not surprised that sufficiently large diffusion will always 
give rise to disorder and
suppress pattern formation. It is interesting though that a medium level of diffusion may cause
pattern formation where the system would otherwise be disordered. 
Let us focus on the example with $P_{11} = -2$, $P_{12} = 2$,  $P_{21} = -2$, 
and $P_{22} = 1$, where subpopulation 2 is repelled from subpopulation 1 and where the self-interaction
within subpopulation 1 is repulsive as well, while the other interactions are attractive. 
According to conditions (\ref{cond1}) and  (\ref{cond2}),
we expect disorder for small fluctuation strengths
$p_1$, $p_2$. While increasing values of $p_2$ should be counterproductive,
increasing $p_1$ should be able to produce pattern formation for
medium values of $p_1$. 
This {\em fluctuation-induced self-organization} is, in fact, observed (see Fig.~\ref{Fig2}a). 
The entities in subpopulation 2 can agglomerate at sites, where the fluctuations 
have temporarily reduced the density in subpopulation 1 due to
disturbances of its homogeneous distribution. 
Later on, subpopulation 1 develops a slightly higher concentration at sites
where subpopulation 2 clusters. In a similar way,
we can have {\em drift-induced self-organization} for $V_1^0 = V > 0$ and $V_2^0 = 0$
(see Fig.~\ref{Fig2}b). For $D_1 = D_2 = 0$, the instability condition (\ref{cond2}) is then replaced by 
\begin{equation}
 \rho_1^0\rho_2^0 (P_{12}P_{21}-P_{11}P_{22}) k^2 (\rho_1^0P_{11}  + \rho_2^0P_{22})^2
 > \rho_1^0\rho_2^0 P_{11}P_{22}(V_1^0-V_2^0)^2 \, ,
\label{cond3}
\end{equation}
where $k$ represents the wave number. (Note that the wave length, 
which is inversely proportional to the respective wave number, is restricted to the
values $\lambda = i \Delta x$ in our cellular automaton simulations.)
\par
\begin{figure}[htbp]
\begin{center}
\begin{picture}(16,14)(0,6.2)
\put(0,21.5){\epsfig{width=7.5\unitlength, angle=-90,
      file=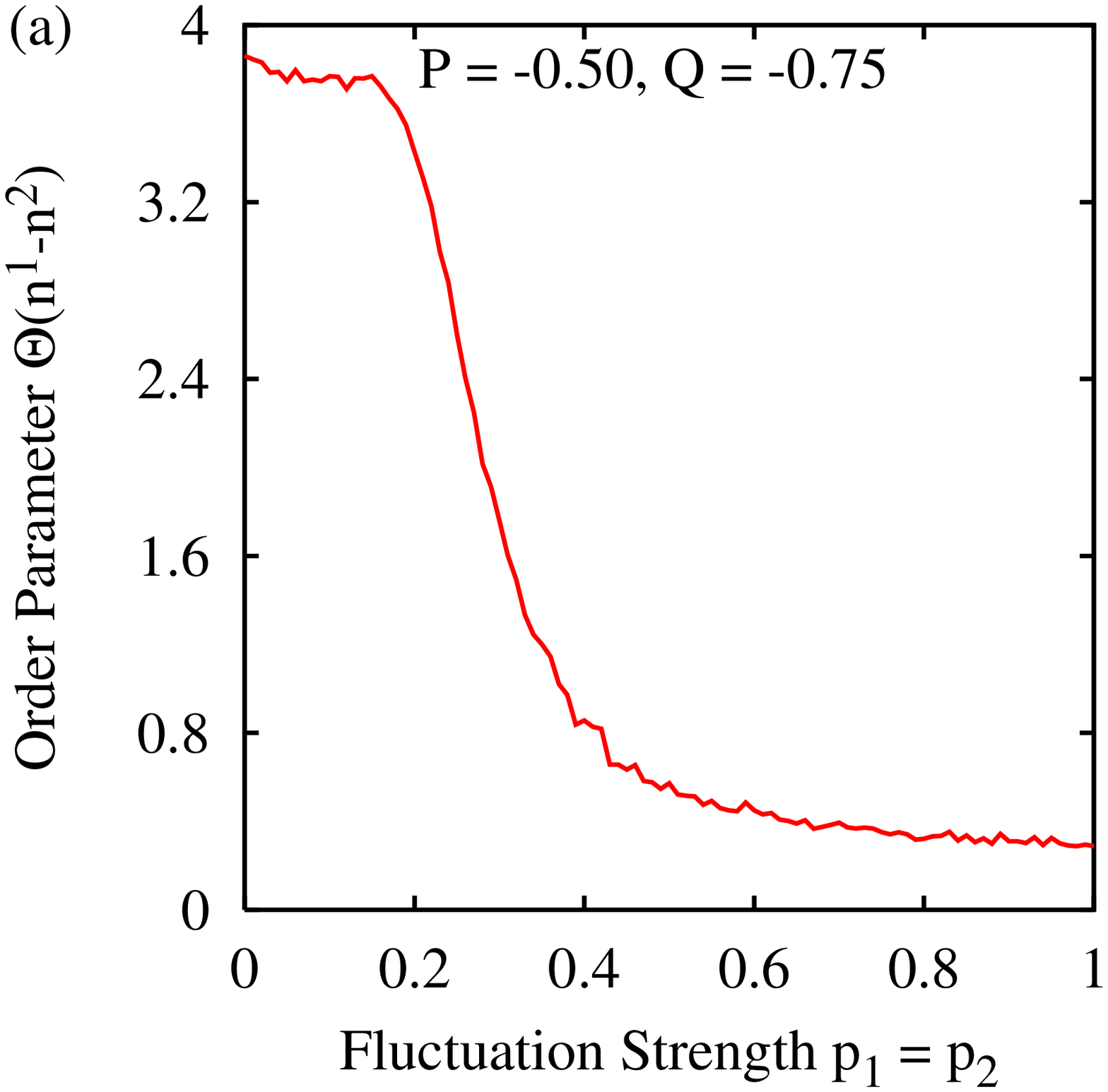}}
\put(7.8,21.5){\epsfig{width=7.5\unitlength, angle=-90,
      file=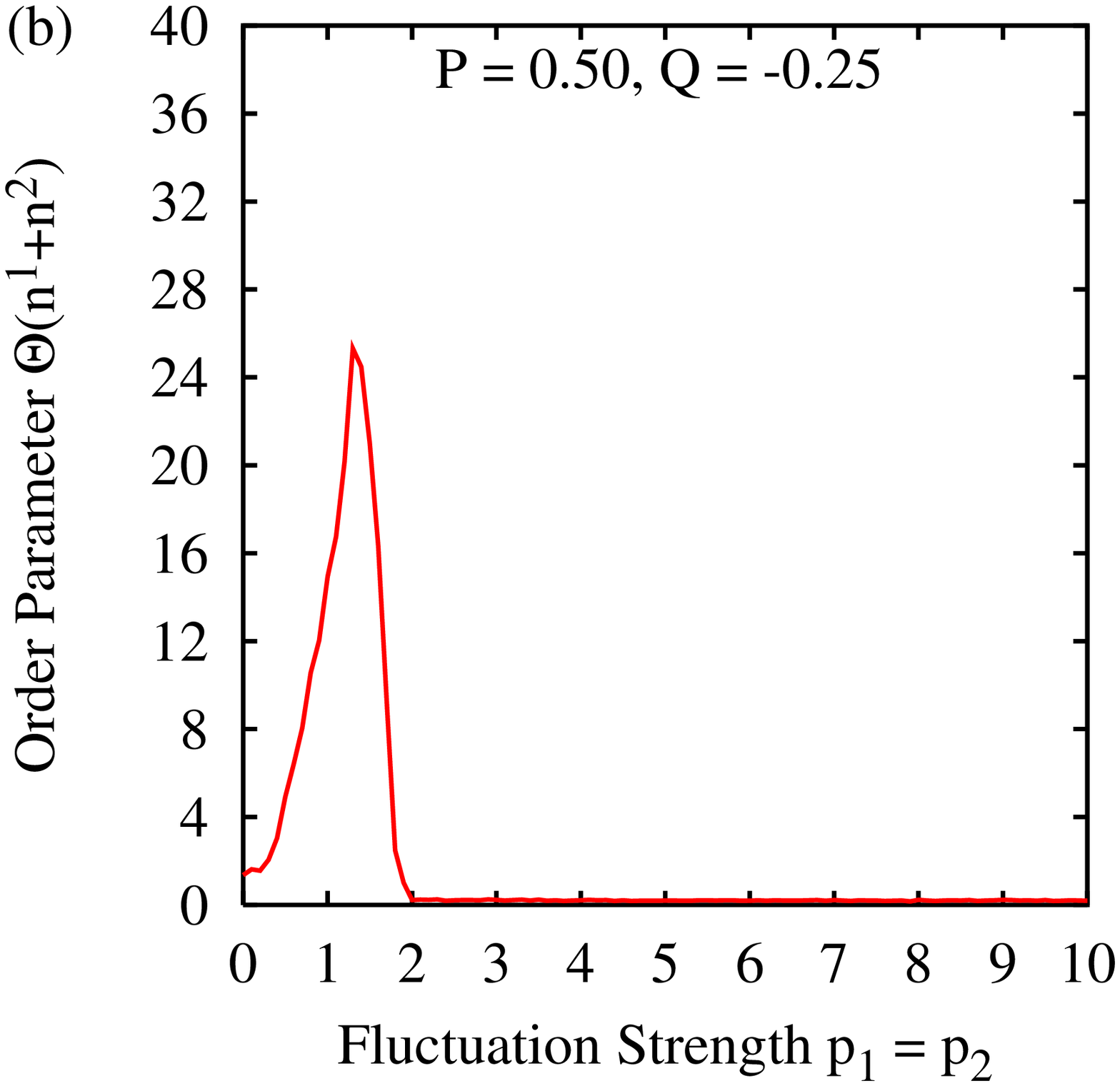}} 
\put(0,13.8){\epsfig{width=7.5\unitlength, angle=-90,
      file=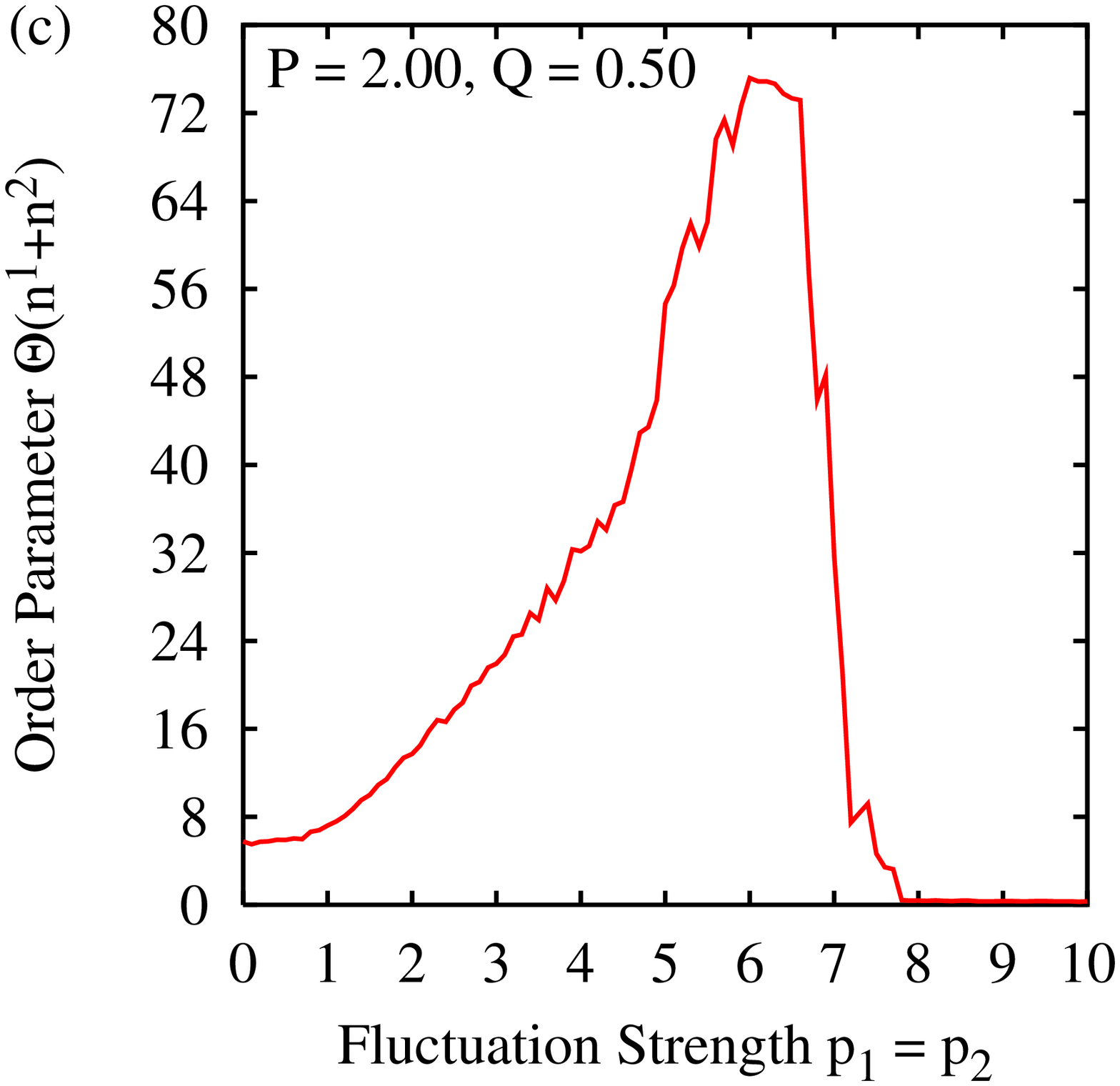}} 
\put(8,13.8){\epsfig{width=7.5\unitlength, angle=-90,
      file=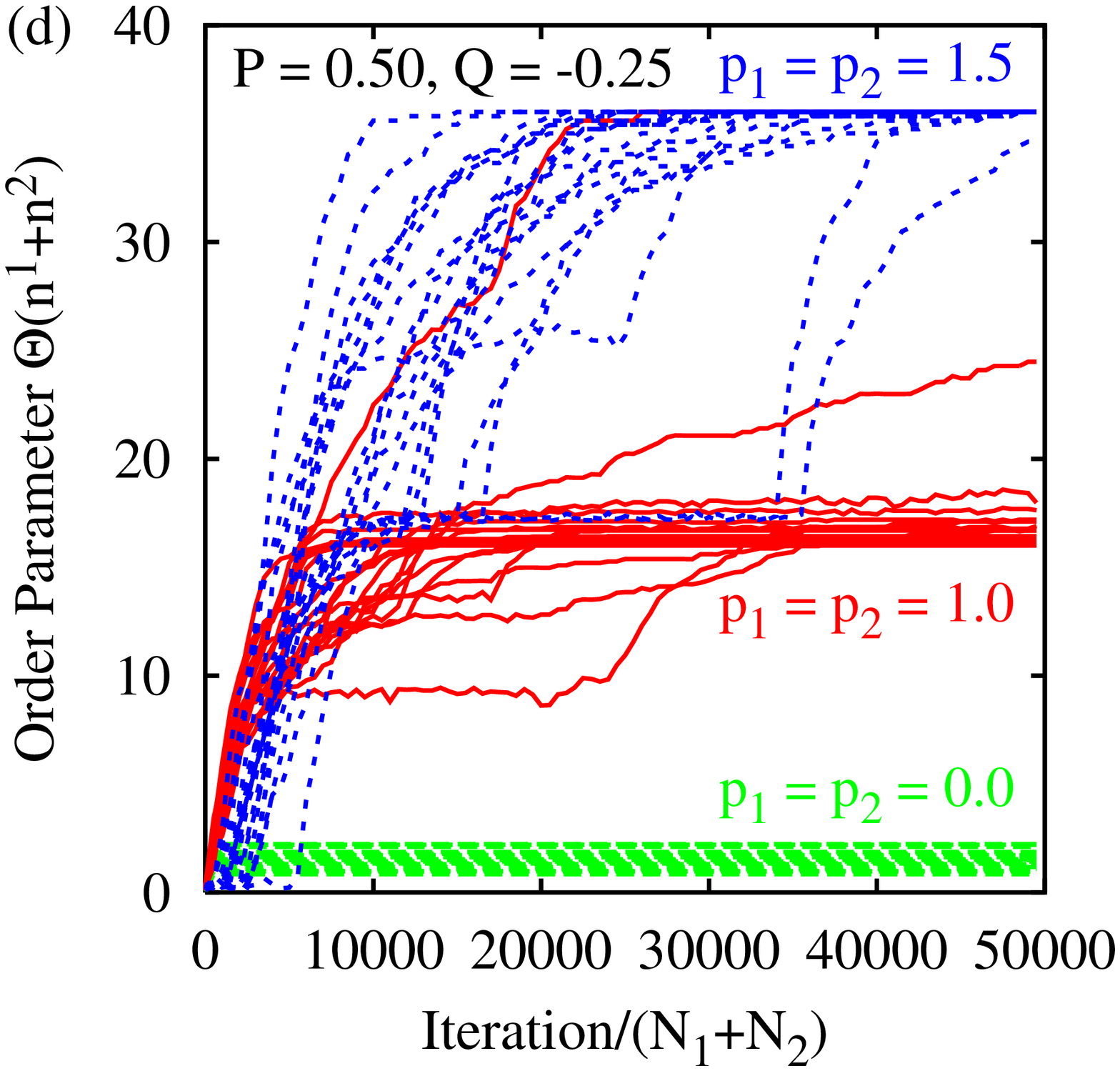}} 
\end{picture}
\end{center}
\caption[]{As Fig.~\ref{Fig2}, but for $V_1^0 = V_2^0 = 0$ and 
symmetric interactions (see the black circles in Fig.~\ref{Fig1}a
for the location in the parameter space). 
(a) In cases of segregation, the order stays about constant
for small noise amplitudes, but it drops significantly for large ones. 
In cases of (b) repulsive or (c) attractive agglomeration,
a suitable noise amplitude $p_1 = p_2$ can increase
the order in the system by more than a factor of 15. The maximum order
is reached at $p_1 = p_2 \approx \max(3 P, 0)$, corresponding to $D_1 = D_2 = \max(\rho P, 0)$.
Close to this maximum, the curves are roughly symmetric, but for less than twice 
this value, the order breaks down completely. (d)  The time-dependence of the
order parameter visualizes the increase of the order
in the system. The plot shows 20 runs for each displayed fluctuation strength, which significantly
influences the dynamics of the ordering process. After large enough times, we find a typical,
noise-dependent length scale and level of order in the system. 
\label{Fig3}}
\end{figure}

For symmetric cases with $P_{11}=P_{22} = P \ne 0$, $P_{12}=P_{21} = Q$, and
$\rho_1^0 = \rho_2^0 = \rho \ne 0$, we find $Q^2 - P^2 > (V_1^0 - V_2^0)^2 / (2 k \rho)^2$,
i.e. a finite drift will usually increase the threshold for pattern formation. This is different
from the effect of diffusion. For the symmetric case with $V_1^0 = V_2^0 = 0$ and
$D_1 = D_2 = D$, the instability condition (\ref{cond2}) reads $\rho^2 Q^2 > (\rho P - D)^2$.
That is, we expect a ``maximum degree of self-organization'' for $D = \max(\rho P,0)$, and 
a more or less symmetric behavior around this point. What does this actually mean?
Fig.~\ref{Fig3} suggests that this statement applies to the order in the system. That is,
for increasing fluctuations strength we find {\em noise-induced ordering} up to 
$D = \max(\rho P,0)$, while we have disorder at significantly higher fluctuations strengths.
\par
One may think that this fluctuation-induced ordering is due to 
noise-induced transitions from a metastable state (local optimum) to a more stable 
state of higher order (possibly the global optimum). We check this hypothesis for
the case of repulsive agglomeration, where a coarse-graining appears particularly 
difficult because of the repulsion effect. Our observations are as follows:
(i) At moderate, but sufficiently large fluctuation strengths, we sometimes
observe a ``step-wise'' fusion of agglomerations, which is associated with exponential-like
relaxation processes of the time-dependent order parameter to a higher level. 
(ii) However, the main mechanism seems to be that, from the very beginning,
fluctuations further the formation of larger agglomerations (i.e. suppress the
development of small ones), which slows down the ordering
process in the early stage. (Sometimes the system stays disordered  for more than
$500\,000(N_1+N_2)$ iterations and, suddenly, the order increases rapidly to a
high level). (iii) A careful choice of the noise strength can speed up the time-dependent 
increase of the order very much. (iv) After a given, large enough
time period, the system has reached a typical level of order, which depends significantly
on the fluctuation strength. In conclusion, one may influence the
typical length scale in the system by variation of the noise level. 
\par
A variation of the ``applied'' drift velocity (if possible) or of the fluctuation strength  
together with a proper choice of the ``treatment times'' would allow one to control pattern formation
in several respects: (i) the speed of ordering, (ii) the typical length scale in the system, 
and (iii) the
level of ordering. A time-dependent variation of the control parameters should even facilitate 
to switch between supporting and suppressing structure formation,
e.g.\ between demixing and homogenization. In the future,
these points may, for example, be relevant for the production, properties, hand\-ling, and transport of
heterogeneous materials, 
for flow control \cite{inject}, and efficient separation
techniques for different kinds of particles. Due to the general, game-theoretical formulation
of the above cellular automaton model, its properties are reminiscent of phenomena
in physical, biological and socio-economic systems. For example, we mention 
phenomena such as the formation of pedestrian lanes (segregation)
\cite{HelMo295,helvic}, of ghettos in cities 
(repulsive agglomeration), or of settlements (attractive agglomeration) \cite{Weidlich}.\\[4mm]

{\em Acknowledgments:} The authors are grateful for financial support by 
the AL\-TA\-NA-Quandt foundation and useful comments by Lutz Schimansky-Geier.
D.H. thanks Ludwig Streit for his invitation and the warm hospitality at the
CCM, Frank Schweitzer for inspiring discussions,
and Tilo Grigat for preparing illustration 1a. T.P. was partially supported 
by KPN Grant 5~P03A~02520.

\end{document}